# A Parallel Quantum Computer Simulator[‡]

Kevin M. Obenland and Alvin M. Despain

*Information Sciences Institute / University of Southern California*

## Abstract

*A Quantum Computer is a new type of computer which can efficiently solve complex problems such as prime factorization. A quantum computer threatens the security of public key encryption systems because these systems rely on the fact that prime factorization is computationally difficult. Errors limit the effectiveness of quantum computers. Because of the exponential nature of quantum computers, simulating the effect of errors on them requires a vast amount of processing and memory resources. In this paper we describe a parallel simulator which accesses the feasibility of quantum computers. We also derive and validate an analytical model of execution time for the simulator, which shows that parallel quantum computer simulation is very scalable.*

# 1.0 Introduction

A quantum computer consists of atomic particles which obey the laws of quantum mechanics. The complexity of a quantum system is exponential with respect to the number of particles. Performing computation using these quantum particles results in an exponential amount of calculation in a polynomial amount of space and time [Feyn85][Beni82][Deut85]. This *quantum parallelism* is only applicable in a limited domain. Prime factorization is one such problem which can make effective use of quantum parallelism[Shor94]. This is an important problem because the security of the RSA public-key cryptosystem relies on the fact that prime factorization is computationally difficult[RiSA78].

Errors limit the effectiveness of any physical realization of a quantum computer. These errors can accumulate over time and render the calculation useless[ObDe96a]. The simulation of quantum circuits is a useful tool for studying the feasibility of quantum computers [ObDe96b]. Simulations inject errors at each step of the calculation and can track their accumulation.

Because of the exponential behavior of quantum systems, simulating them on conventional computers requires an exponential amount of operations and storage. For this reason, to simulate problems of interesting size we must employ the use of parallel supercomputers. In this paper we describe a parallel simulator which models the accumulation of errors in a quantum computer. We implemented the simulator in 'C' and use the MPI message passing library to communicate between processors [MPI94]. The parallel simulator allows the simulation of circuits which are three to four orders of magnitude larger than any current proposed experimental realizations of a quantum computer[TuHo95][MoMe95].

The remainder of this section gives a brief introduction to quantum computers. In section 2 we describe our simulation methodology and in section 3 we describe the parallel simulator. In section 4 we use an analytical model of execution time to evaluate the scalability of the parallel simulator and in section 5 we conclude.

## 1.1 Quantum Computers

A quantum computer performs operations on bits, called *QU-bits*, whose values can take on the value of one or zero or a superposition of one and zero. This superposition allows the representation of an exponential number of states using a polynomial number of QU-bits. A quantum computer performs transformations on these QU-bits to implement logic gates. Combinations of these logic gates define quantum circuits.

## 1.2 QU-bits and Quantum Superposition

The basic unit of storage in a Quantum Computer is the *QU-bit*. A QU-bit is like a classical bit in that it can be in two states, zero or one. The QU-bit differs from the classical bit in that, because of the properties of quantum mechanics, it can be in both these states simultaneously[FeLS65]. A QU-bit which contains both the zero and one values is said to be in the *superposition* of the zero and one states. The superposition state persists until we perform an external measurement. This mea-surement operation forces the state to one of the two values. Because the measurement determines without doubt the value of the QU-bit, we must describe the possible states which exist before the measurement in terms of their probability of occurrence. These QU-bit probabilities must always sum to one because they represent all possible values for the QU-bit.

| Bit Value | Amplitude | Probability |
|---|---|---|
| 0 | 1 | 1 |
| 1 | 0 | 0 |

**(a)** Representation of a 0 bit value

| Bit Value | Amplitude | Probability |
|---|---|---|
| 0 | $1/(\sqrt{2})$ | $1/2$ |
| 1 | $1/(\sqrt{2})$ | $1/2$ |

**(b)** Representation of a superposition between 0 and 1

**FIGURE 1. Vector representation of QU-bit values**

The quantum simulator represents a QU-bit using a state vector. Figure 1 shows how the simulator uses complex amplitudes to represent a QU-bit. Each state in the vector represents one of the possible values for the QU-bit. The bit value of a state corresponds to the index of that state in the vector. The simulator represents each encoded bit value with a non zero amplitude in the state vector. The probability of each state is defined as the square of this complex amplitude[FeLS65]. Figure 1(a) shows a state which represents the single value of zero. In Figure 1(b) the probability is equally split between the zero and one states, representing a QU-bit which is in the superposition state. For a register with M QU-bits, the simulator uses a vector with $2^M$ states.

An M QU-bit register can represent $2^M$ simultaneous values by putting each of the bits into the superposition state. A calculation using this register calculates all possible outcomes for the $2^M$ input values, thereby giving exponential parallelism. The bad news is that in order to read out the results of a calculation we have to observe, i.e. measure, the output. This measurement forces all the bits to a particular value thereby destroying the parallel state. The challenge then is to devise a quantum calculation where we can accumulate the parallel state in non-exponential time before performing the measurement.

## 1.3 Quantum Transformations and Logic Gates

A quantum computation is a sequence of transformations performed on the QU-bits contained in quantum registers[Toff81][FrTo82][Feyn85][BaBe95][Divi95]. A transformation takes an input quantum state and produces a modified output quantum state. Typically we define transformations at the gate level, i.e. transformations which perform logic functions. The simulator performs each transformation by multiplying the $2^M$ dimensional vector by a $2^M$ x $2^M$ transformation matrix.

The basic gate used in quantum computation is the *controlled-not*, i.e. exclusive or gate. The controlled-not gate is a two bit operation between a *control* bit and a *resultant* bit. The operation of the gate leaves the control bit unchanged, but conditionally flips the resultant bit based on the value of the control bit. Table 1 shows a truth table of how the controlled-not gate modifies the different QU-bit values. In the vector representation of the QU-bits, the controlled-not gate corresponds to a transformation which swaps the amplitude of the states in the third and fourth positions. Figure 2 shows the four by four matrix which performs the controlled-not transformation on the two QU-bits.

**TABLE 1. Truth Table for the Controlled-not gate**

| Input Bits | | Output Bits | |
|---|---|---|---|
| A | B | A' | B' |
| 0 | 0 | 0 | 0 |
| 0 | 1 | 0 | 1 |
| 1 | 0 | 1 | 1 |
| 1 | 1 | 1 | 0 |

$$\begin{bmatrix} 1 & 0 & 0 & 0 \\ 0 & 1 & 0 & 0 \\ 0 & 0 & 0 & 1 \\ 0 & 0 & 1 & 0 \end{bmatrix} \times \begin{bmatrix} a_0 \\ a_1 \\ a_2 \\ a_3 \end{bmatrix} = \begin{bmatrix} a_0 \\ a_1 \\ a_3 \\ a_2 \end{bmatrix} \quad \leftarrow \text{Swap Amplitudes } a_2 \text{ and } a_3$$

**FIGURE 2. Controlled not Transformation**

### 1.4 The Physical Realization of a Quantum Computer

The ion trap quantum computer, proposed by Cirac and Zoller, is one of the most promising schemes for the experimental realization of a quantum computer[CiZo95]. Other promising schemes include cavity QED[LaTu95][TuHo95] and Quantum dots[BaDe95]. To date, cavities trapping up to 33 QU-bits have been constructed[RaGi92], and simple quantum gates have been demonstrated[MoMe95]. We model our simulations directly on the Cirac and Zoller scheme. The ion trap computer uses the internal energy states of the ions to encode the zero and one states of a QU-bit. This scheme also requires a third temporary state to implement gates such as the controlled-not gate. Lasers directed at the individual ions define a series of transformations which when combined implement the controlled-not gate.

In general the physical process used to perform the transformations will not be performed perfectly. The resulting inaccuracies, referred to as *operational* errors, degrade the calculation over time. Also interaction of the QU-bits with the external environment has a destructive effect on the coherence of the superposition state. This type of error, referred to as *decoherence*, acts to measure the QU-bits and thereby destroys the parallel state. Both operational and decoherence errors act to limit the effectiveness of a quantum computer. Simulation is an effective tool for characterizing errors, and tracking their accumulation. Using a physical model such as the Cirac and Zoller scheme is important for obtaining realistic results.

## 1.5 Quantum Algorithms

Much of the current interest in quantum computation is due to the discovery of an efficient algorithm to factor numbers[Shor94]. This is an important problem because a quantum factoring engine would severely threaten the security of public-key cryptosystems. The quantum factoring algorithm uses quantum parallelism to calculate all of the values of a function simultaneously. This function is periodic and a quantum FFT can extract this period efficiently [Copp94]. We then use a polynomial time classical algorithm to calculate the factors from this period.

Quantum computers are also useful for searching unsorted databases[Grov96]. The quantum search algorithm runs in $O(\sqrt{N})$ time for $N$ items, where the best classical algorithm runs in $O(N)$ time. Therefore for NP-complete problems such as circuit SAT, which contains $2^m$ items for $m$ variables, the quantum algorithm runs in $O(2^{m/2})$ time. This speedup is not exponential like that of the factoring algorithm, but it allows the solution of problems that may be computationally intractable on a classical computer.

## 2.0 Quantum Simulation Methodology

We implemented the quantum simulator entirely in the 'C' programming language. The simulator takes as input the description of a quantum circuit specified in terms of logic gates. The simulator implements one, two and three bit controlled-not gates as well as rotation gates. The simulator implements gates as sequences of laser transformations as defined by the Cirac and Zoller trapped ion scheme. For example, to implement a controlled-not gate, the simulator uses a sequence of five laser transformations. Each of these laser operations are essentially rotations between the zero and one values of a QU-bit.

Imperfections in the laser apparatus result in deviations in the angle of rotation. The simulator operates at two different levels of detail. We base the most detailed model directly on the Cirac and Zoller scheme. This model requires extra memory and processing resources to model the third temporary state. To represent M QU-bits, we now require $3^M$ states, and transformations are now matrix multiplications using matrices of size $3^M \times 3^M$. A less detailed model, which we have shown is very accurate, uses only two states for each QU-bit and therefore represents an exponential reduction in complexity [ObDe97].

## 2.1 Simulation of Quantum Logic Gates

The simulator implements each gate as a sequence of transformations on the vector space which represents the QU-bits in the computer. In general, for a quantum computer with a vector space of size V, we describe each operation with a VxV transformation matrix. Because each gate only operates on a small number of bits, these matrices are simply 2x2 or 4x4 matrices replicated many times. We can avoid having to represent the entire matrix by iterating over the vector space and performing the simpler operations for all of the corresponding sets of states. Figure 3 shows an example of how the simulator performs a 2x2 transformation on a four element state vector. The simulator splits the vector into separate two elements vectors and performs independent 2x2 transforms on each of the pieces. The pieces when combined form a new four element vector. This procedure applies to vectors of any size, where a 2x2 transformation on a vector of size V, is a sequence of V/2 independent 2x2 transformations. To perform a transformation on a particular bit, the simulator splits the state space so that the bit value of the states in the vector differ only in the position of that bit.

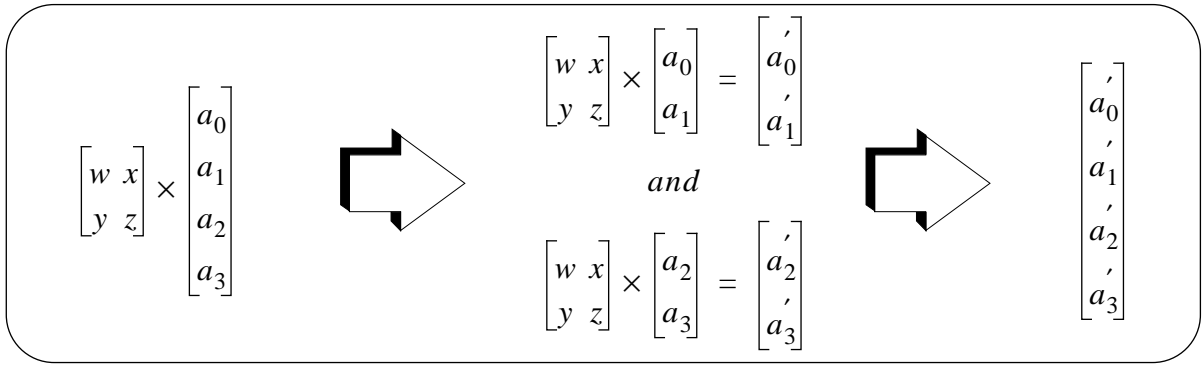

**FIGURE 3. A 2x2 transformation performed on a 4 state vector**

## 2.2 Quantum Circuits

Our simulation studies use circuits of various size which all implement Shor's factoring algorithm. The factoring algorithm works by using quantum parallelism to compute all the values of a function simultaneously. By putting the QU-bit register *A* in the superposition of all values and calculating the function $f(A) = X^A mod\ N$, we calculate all the values of *f(A)* simultaneously. Where N is the number we want to factor, and *X* is a randomly selected number which is relatively prime to N. A schematic of a circuit for factoring the number 15 is shown in Figure 5, where we use rotation gates to prepare the superposition state. The function f(A) is periodic, and we can extract this period by performing a quantum FFT on the amplitudes in the *A* register[Copp94]. Figure 4 shows a plot of the function f(A) where X=7 and N=15. As the figure shows, the function repeats with a period of four.

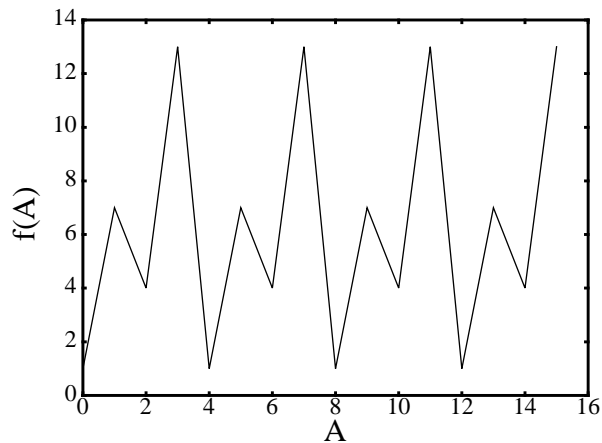

**FIGURE 4. Function $f(A) = 7^A$ mod 15**

As we see from Figure 5 each quantum factoring circuit consists of three pieces: (1) preparing the superposition state, (2) calculating the function *f(A)* and (3) performing the quantum FFT. We do not include the portion of the calculation which extracts the period because we perform it off-line using a classical computer. The circuit to calculate *f(A)* represents the majority of the calculation in the quantum circuit.

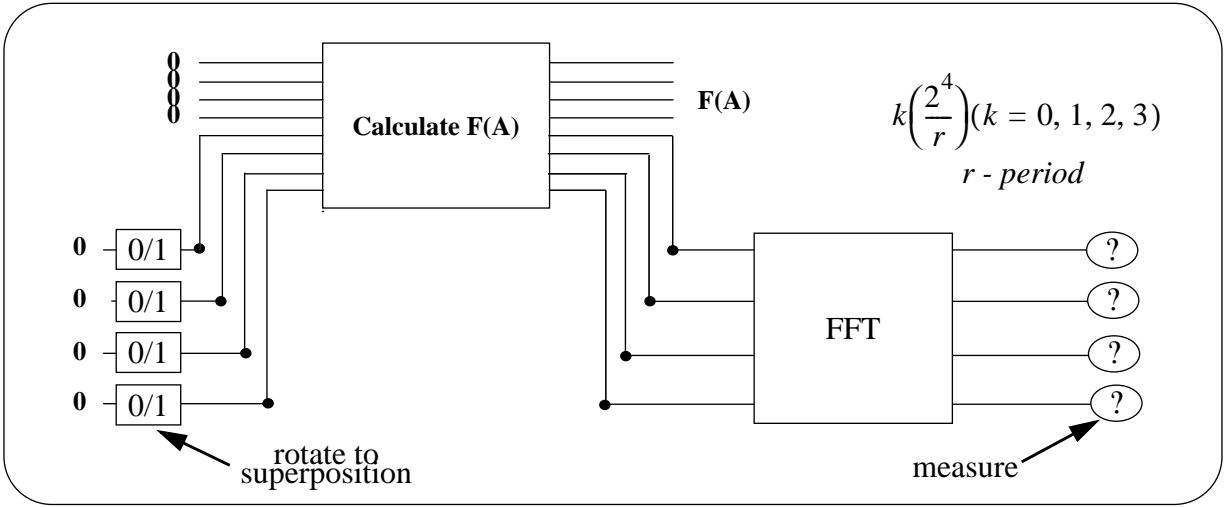

**FIGURE 5. Factor 15 Circuit used in simultaneous**

### 2.2.1 Factoring Implementation

We use repeated squaring to perform the modulo exponentiation required by the *f(A)* circuit[Desp96][BeCh96]. Equation 1 shows that if we use the binary representation of *A,* the exponentiation consists of a sequence of multiplications. At each step we use a single bit of A and multiply a running product by the factor $X^{2^l}$. Each of these multiplications requires time $O(L^2)$, where L is the number of bits required to represent the number to factor, N. In order to extract the period from the quantum calculation, Shor suggests using 2L + 1 bits for the input register A[Shor94], but we have shown that an accurate simulation can be achieved using only L + 1 bits for A [ObDe97].

$$f(A) = X^A \bmod N = X^{[a_0 2^0 + a_1 2^1 \ldots a_{l-1} 2^{l-1}]} \bmod N = X^{a_0 2^0} \bullet X^{a_1 2^1} \ldots X^{a_{l-1} 2^{l-1}} \bmod N \quad \text{(EQ 1)}$$

The circuit also requires two additional scratch registers to write the intermediate results of the additions and multiplications. The number of QU-bits required to factor an L bit number using this circuit is therefore, 4L + 4. The total number of operations using this method is $252L^3 + 8L^2 + L + 3$. Because all additional processing needed to extract the factor can also be done in polynomial time, Shor's algorithm gives an efficient means of factoring large numbers.

# 3.0 Parallel Simulation

Because of the exponential nature of quantum computing, simulation of it on a conventional computer requires exponential memory and processing resources. For a medium sized problem, like a single modulo multiplication step from the Factor 15 problem, we need to represent of $3^{16} =$ 43,046,721 states if we use the detailed three state model. This translates to over 600 Megabytes of storage. There are also about 8000 operations in the simulation, and each operation must operate on each of the states. Fortunately we can split up the calculation and use the memory and processing resources of multiple processors.

## 3.1 Parallel Division of a Quantum Computer Simulation

As shown in Figure 3, quantum simulation involves a series of independent transformations on the vector space used to represent the QU-bits. Because of this inherent parallelism, quantum computer simulation is a natural candidate for parallel processing. For an M QU-bit quantum computer, the vector space for the detailed Cirac and Zoller model is of size $V=3^M$. A single transformation on this state space is therefore a sequence of $3^{M-1}$ simple operations which can all be run in parallel. For a circuit with 16 QU-bits, this represents a maximum degree of parallelism of $3^{15}$.

One potential obstacle to parallelizing a quantum computation is the reorganization of data that may be needed between the simulation of successive gates. The bits used by a gate define how the simulator divides the vector space into two element vectors. To exploit all the available parallelism, the simulator must partition all the two element vectors to separate processors. Because there is excess parallelism, the simulator can allocate multiple two element vectors on each processor. The allocation needs to change only when we need to perform a transformation which is not covered by the original allocation.

To exploit a degree of parallelism of $3^N$, the simulator picks a set of N QU-bits to parallelize across. To determine the allocation of the QU-bit states, the simulator concatenates the N QU-bit values of each state and allocates that state to the processor whose ID is equal to the concatenated value. For a quantum simulation of M QU-bits, each of the processors will have $3^{M-N}$ states allocated on it. All of the processors can run in parallel operating only on local data as long as they do not use any of the QU-bits in the set N. When the computation needs to operate on one of the bits in N, a new set is picked and the simulator redistributes the data. The entire simulation is a sequence of these computation and reorganization steps. The efficiency of the parallel simulator now becomes a question of how often the simulator needs to redistribute the data.

Quantum factoring is a good candidate for parallel simulation because it consists of large sections of operations which operate on a subset of the total QU-bits. As shown in Equation 1, the calculation of *f(A)* consists of a set of multiplications each using only a single bit of *A*. The simulator can therefore exploit parallelism across the unused bits of A. The simulator can also exploit parallelism within a multiplication because each multiplication is a sequence of additions. Figure 6 shows in pseudo code the way in which we perform this decomposition for an L bit number. We calculate *f(A)* as $X^{A\_bit}$ multiplied by the running product P. This product is calculated for each bit of P into the running sum S using modulo addition. The Modulo addition step comprises the majority of the calculation and operates on a single bit of A and P, and therefore the simulator can exploit parallelism across the unused bits of A and P.

*For A_bit = A[0].. A[L]* /* calculate f(A) as a sequence of multiplications */
   *P = 1*   /* P is the running product */
   *For P_bit = P[0] .. P[L-1]*  /* calculate a product as a sequence of additions */
      *S = Modulo Add(S,X,A_bit,P_bit)*  /* Operate on a single bit of A and P */
   *P = S*

**FIGURE 6. Pseudo Code to calculate f(A)**

## 3.2  Dynamic Allocation of the State Vector

The simulator uses a hierarchical linked structure to represent the QU-bit vector space. This structure is very similar to a binary tree, except that the number of elements at each level is configurable. The structure uses zero or more link levels and a single level containing the state values. Figure 7 shows an example of a three level structure. The number of bits which a level covers, determines the size of the block at that level. For the Cirac and Zoller three state model, a block will have $3^B$ elements for a level covering B QU-bits. The simulator allocates a block for a level when it first uses a bit at that level. This allocation causes the simulator to allocate space for all bits covered by that level. Null pointer values at the link level mark blocks which have not been allocated.

One extreme organization, for the linked structure, is to have a level for every QU-bit. This allocation strategy has a fine granularity and therefore the simulator allocates blocks only when needed. The disadvantage is that the simulator must traverse the maximum number of levels to get to the state values.The other extreme is a flat structure without any link values. This has the disadvantage that the simulator must allocate the complete state space at the start of the simulation. The best compromise is to allocate enough link levels so that bits which are not used until later in the simulation are covered by link levels. This assures that the simulator will allocate the state values for these bits only when the bits are used.

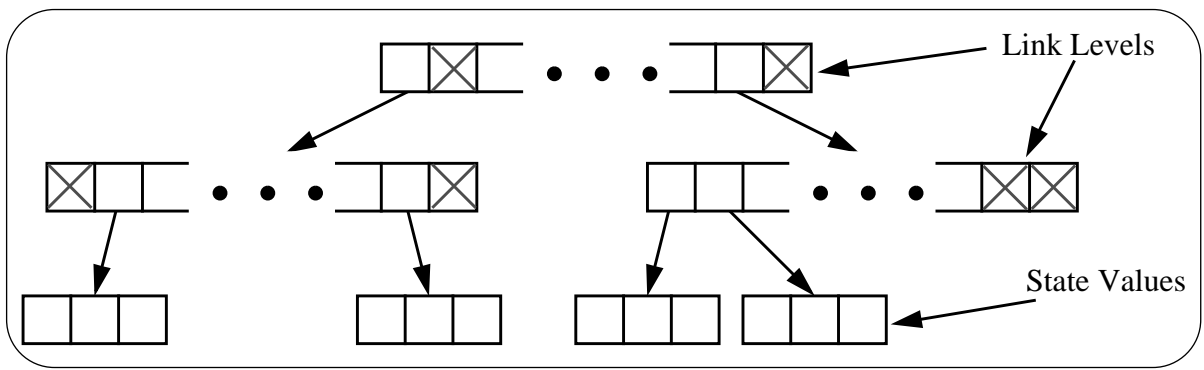

**FIGURE 7. Hierarchical linked state vector structure**

The hierarchical structure has three advantages over a flat structure. The first advantage is that by delaying the allocation of some of the state space, we eliminate unnecessary calculation. This is because all the un-allocated states have an amplitude of zero, and therefore all transformations on them have no effect. The second advantage is that the structure is the same regardless of the number of states used to represent a QU-bit. For example a simulation which uses the detailed three state Cirac and Zoller model has the same structure as a model which uses only two states to represent a QU-bit. The last advantage is that the parallel simulator can reorganize data by redistributing the state value blocks. If the simulator never exploits parallelism across the QU-bits covered by the lowest level, the data in the state value blocks is never split across processors. In a flat structure, to distribute data from one processor to another, the data would not be contiguous and the simulator would have to copy the data to a buffer before sending it. The receiving processor would then have to unpack the data into its' own flat structure.

## 3.3 Parallel Execution

In a parallel simulation each processor performs a portion of the operations required to implement each of the laser transformations. As described in Section 3.1 the simulator picks a set of bits to parallelize across, and then redistributes the data. Each processor allocates a duplicate copy of all the link blocks in the hierarchical vector structure. A processor however only allocates space for the state values assigned to it.

To perform the operations in a transformation, each processor traverses the linked state vector and performs the calculations on its local states. When all the processors have finished the computation phase, they reorganize the state vector by exchanging the state value blocks. Figure 8 shows the algorithm for redistributing the data between the current organization, defined by the variable *current_parallel_bits*, and a new organization, defined by the variable *new_parallel_bits*.

```
For Block_num = 0 .. Num_value_blocks - 1
    current_proc_num = concatenate_parallel_bits(Block_num,current_parallel_bits)
    new_proc_num = concatenate_parallel_bits(Block_num,new_parallel_bits)
    if ( current_proc_num == new_proc_num )
        /* do nothing, the data is to remain on the same processor */
    else if ( current_proc_num == my_pid )   /* this processor currently owns the data */
        send_data_block(new_proc_num)  /* send to the new owner */
    else   /* this processor is to own the data for the next step */
        receive_data_block()  /* receive the data from the current owner */
```

**FIGURE 8. Redistributing the state data**

## 3.4 Scalability of Parallel Simulation

As discussed in Section 3.1, simulation of quantum factoring has a large amount of potential parallelism. We will use the six factoring benchmarks shown in Table 2 to evaluate the parallel simulator. The Modulo Multiply benchmark is a portion of the factor-15 circuit. It is the largest circuit that the simulator can feasibly simulate using the full three state model. There are two factor-15 benchmarks, one which uses 9 bits in the A register, and one which uses only 3. We have shown that the three bit factor-15 circuit is a good approximation of the 9 bit circuit [ObDe97]. The factor-21 and factor-35 benchmarks factor five and six bit numbers respectively.

The total number of basic operations needed in a simulation is proportional to the number of states multiplied by the number of laser operations. The actual number of operations will be less than the maximum because the use of the dynamic allocated state vector eliminates some of the operations. To simulate decoherence errors the simulator performs two additional transformations after every laser operation, increasing the total number of transformations by a factor of three.

**TABLE 2. Factoring benchmarks**

| Benchmark | Description | Number of States | Number of Laser Operations | Number of Link Bits | Number of State Bits |
|---|---|---|---|---|---|
| Mult2 | Modulo Multiply, two state | $2^{16}$ | 7,137 | 10 | 6 |
| Mult3 | Modulo Multiply, three state | $3^{16}$ | 8,854 | 10 | 6 |
| f15_9bits | Factor 15, 9 A bits | $2^{24}$ | 70,904 | 16 | 8 |
| f15_3bits | Factor 15, 3 A bits | $2^{18}$ | 70,793 | 12 | 6 |
| f21 | Factor 21 | $2^{24}$ | 139,678 | 16 | 8 |
| f35 | Factor 35 | $2^{28}$ | 237,798 | 20 | 8 |

For an N QU-bit simulation the parallel simulator can exploit parallelism across up to N-1 bits. In general this is too much parallelism for even the largest supercomputer, so the simulator will exploit parallelism across a smaller set of bits. The number of reorganization steps also limits the effectiveness of parallelizing the simulation. Table 3 shows the number of reorganization steps, as a function of the number of parallel bits, required for each benchmark assuming the number of link bits given in Table 2.

Because the state value blocks can not be split between processors, the simulator can only exploit parallelism across bits allocated at the link level. Also because the dynamic structure eliminates the calculation for unused bits, the simulator must parallelize across bits which have been previously used. Because of these factors the simulator cannot always exploit the maximum amount of parallelism. By increasing the number of link bits, the simulator can increase the amount of parallelism, but this may increase the overhead of traversing the link data structure.

For the Mult3 benchmark the amount of parallelism is $3^N$ for N parallel bits, and for all other benchmarks the amount of parallelism is $2^N$. As Table 3 shows the number of reorganization steps is very low compared to the total number of operations for any number of parallel bits. There is also more than enough parallelism for all benchmarks. For example, if the Mult3 benchmark is parallelized across four bits, it can be run on $3^4$=81 processors and still require only 46 reorganization steps.

**TABLE 3. Number of Reorganization Operations**

| Benchmark | Number of Parallel Bits | | | | | | | | | | |
|---|---|---|---|---|---|---|---|---|---|---|---|
| | 1 | 2 | 3 | 4 | 5 | 6 | 7 | 8 | 9 | 10 | 11 |
| Mult2 | 6 | 7 | 22 | 46 | N/A | N/A | N/A | N/A | N/A | N/A | N/A |
| Mult3 | 6 | 7 | 22 | 46 | N/A | N/A | N/A | N/A | N/A | N/A | N/A |
| f15_9bits | 4 | 8 | 21 | 71 | 161 | 363 | 582 | N/A | N/A | N/A | N/A |
| f15_3bits | 4 | 8 | 28 | 46 | 100 | 321 | 598 | 1025 | 2570 | N/A | N/A |
| f21 | 3 | 7 | 18 | 36 | 90 | 184 | 319 | 474 | 905 | 1583 | 2127 |
| f35 | 3 | 7 | 17 | 31 | 48 | 75 | 167 | 300 | 479 | 601 | 868 |

## 3.5 Analytical Model of Execution Time

In this section we derive an analytical model of execution time to evaluate the parallel simulator. The model calculates the number of operations in a simulation and uses this to determine the expected execution time. Table 4 describes the parameters used in all the equations.

**TABLE 4. Parameters used in the analytical model**

| Parameter | Description |
|---|---|
| s | Number of states used to represent a QU-bit (two or three) |
| $t_t$ | Time to traverse the link structure (In units of seconds/number of link bits) |
| $t_{lat}$ | Latency of sending a message (In units of seconds.) |
| $t_b$ | Cycle time required to transfer the data (In units of seconds/byte) |
| $t_{op}$ | Time to perform an operation on a single state (In seconds/states) |
| $t_{lr}$ | Time required for each reorganization message (In seconds/data block) |
| q | Size of a QU-bit state (In bytes) |
| $l_{total}, l_{step}$ | Number of laser operations (In the total simulation or for a sequence of steps) |
| $n_b$ | Total number of bits used for a sequence of operations |
| $n_l$ | Number of bits in the link level for a sequence of operations |
| $n_p$ | Number of parallel bits |
| $n_c$ | Number of new parallel bits for a reorganization step |

Equation 2 defines the parameterized function $T_{reorg}$, which computes the time required to perform a single reorganization step. The time equation consists of two parts, the number of data blocks multiplied by the time to reorganize a single data block. The number of link bits determines the total number of data blocks $s^{n_l}$. If some of the parallel QU-bits remain after the reorganization step, then the number blocks which need to be reorganized is reduced by the factor $s^{n_l - n_c}$. The reorganization time for each block consists of the latency of sending a message, the amount of time required to transfer the message and the computation time required to process the message.

$$T_{reorg}(n_l, n_b, n_c) = (s^{n_l} - s^{n_l - n_c})(t_{lat} + t_b q s^{n_b - n_l} + t_{lr}) \qquad \text{(EQ 2)}$$

Equation 3 defines $T_{dec}$ which gives the extra communication required for a complete simulation which includes decoherence. The parallel simulator must accumulate and redistribute a sum from all the processors after every laser operation. The decoherence step adds two additional operations for every laser operation, and requires two messages to accumulate and redistribute the sum.

$$T_{dec} = 2 l_{total} t_{lat} \qquad \text{(EQ 3)}$$

Equation 4 defines the parameterized function $T_{comp}$, the amount of time required to simulate a sequence of laser operations. $T_{comp}$ consists of two parts, the time required to traverse the link structure and the time required to perform the transformations on the state data. All other computation is insignificant and is ignored.

$$T_{comp}(l, n_b, n_l) = l(s^{n_l} t_t + s^{n_b - 1} t_{op}) \qquad \text{(EQ 4)}$$

The total time for a flat sequential simulation $T_{seqflat}$, as given by Equation 5, is just the computation time for all the laser operations. There are no link bits and therefore the time to traverse the link structure is zero.

$$T_{seqflat} = T_{comp}(l_{total}, n_{b, total}, 0) \qquad \text{(EQ 5)}$$

$T_{seqdyn}$, The total time for a simulation which uses the dynamic state vector is shown in Equation 6. The dynamic structure grows as bits are used, therefore there are steps in the calculation which have fewer states and require less simulation time. The total simulation time is the sum of these steps.

$$T_{seqdyn} = \sum_{step = 1}^{totalsteps} T_{comp}(l_{step}, n_{b, step}, n_{l, step}) \qquad \text{(EQ 6)}$$

The parallel simulator always uses the dynamic state vector and the simulation time $T_{par}$, as given by Equation 7, is also broken into steps. The parallel simulation time also includes the time required to reorganize the data when the set of parallel bits changes. Parallelizing across $n_p$ QU-bits reduces the effective size of the calculation any processor must perform by a factor of $s^{n_p}$.

$$T_{par} = \sum_{step = 1}^{totalsteps} T_{comp}(l_{step}, n_{b, step} - n_p, n_{l, step}) + T_{reorg}(n_{b, step}, n_{l, step}, n_{c, step}) \qquad \text{(EQ 7)}$$

Equation 8 and Equation 9 show the time required by a simulation which includes decoherence. The simulator must perform two additional transformations after every laser operation, increasing the execution time by a factor of three. For a parallel simulation, the simulator requires extra communication to accumulate a sum during the decoherence transformation.

$$T_{pardec} = 3T_{par} + T_{dec} \qquad \text{(EQ 8)}$$

$$T_{seqdec} = 3T_{seq} \qquad \text{(EQ 9)}$$

### 3.5.1 Measured Values for the Time Parameters

We ran a series of tests on a Cray T3E and an IBM SP2 to get typical values for the time constants used in the analytical model. The Cray T3E is a 256 processor machine which uses a 300 MegaHertz DEC Alpha 21164 CPU with 128 Mega-bytes of memory per processor. The IBM SP2 is a 30 processor machine which uses the Power2 CPU with 128 Mega-bytes of memory per processor. We obtained the timings shown in Figure 9 using single runs of the f15_3bits and Mult3 benchmarks. We performed separate tests for the three state and two state simulations because the two state simulator uses a special transformation to approximate the transformations of a three state simulation.

| Parameter | Value |
| --- | --- |
| $t_{op}$ (2 state) | 1.5µsec |
| $t_{op}$ (3 state) | 900nsec |
| $t_t$ | 450nsec |
| $t_{lat}$ | 6µsec |
| $t_b$ | 1nsec/byte |
| $t_{lr}$ | 80µsec |

(a) T3E

| Parameter | Value |
| --- | --- |
| $t_{op}$ (2 state) | 2.3µsec |
| $t_{op}$ (3 state) | 1.28µsec |
| $t_t$ | 1µsec |
| $t_{lat}$ | 20µsec |
| $t_b$ | 20nsec/byte |
| $t_{lr}$ | 500µsec |

(b) SP2

**FIGURE 9. Measured time values**

### 3.5.2 Validation of the Analytical Model

To validate the analytical model we gathered timings for additional runs of the simulator. We ran all simulations with the size parameters given in Table 2 and with the number of reorganizations given in Table 3. Table 5 compares the observed wall clock time to the calculated time for three benchmarks on the SP2. Because of the small size of the machine the largest number of parallel processors that we could use was nine. As the table shows the calculated time is close to the observed time and is off by at most 8%.

**TABLE 5. Observed and Calculated run times for the SP2**

| Benchmark | Number of Processors | Total Time (seconds) | Calculated Time (seconds) | Relative Difference |
|---|---|---|---|---|
| f15_3bits | 1 | 15620 | 16111 | +3.1% |
|  | 4 | 4467 | 4196 | -6.1% |
|  | 8 | 2319 | 2234 | -3.7% |
| Mult3 | 9 | 12786 | 13809 | +8.0% |
| f15_9bits | 4 | 73518 | 69977 | -4.8% |

Table 6 shows the observed and calculated times for the T3E using four different benchmarks. The T3E has 256 processing elements which allows us to run larger simulations, i.e. up to 128 processors. As Table 6 shows, the calculated times are very close to the observed times for all two state simulations. For the three state simulations, i.e. the Mult3 benchmark, the calculated time for 81 processors is off by just over 10%.

**TABLE 6. Observed and Calculated run times for the T3E**

| Benchmark | Number of Processors | Total Time (seconds) | Calculated Time (seconds) | Difference |
|---|---|---|---|---|
| Mult2 | 1 | 288 | 282 | -2.1% |
|  | 2 | 140 | 143 | +2.1% |
|  | 4 | 72 | 73 | +1.4% |
|  | 8 | 37 | 38 | +2.7% |
| Mult3 | 27 | 3282 | 3334 | +1.6% |
|  | 81 | 1506 | 1345 | -10.7% |
| f15_3bits | 8 | 1402 | 1398 | -0.3% |
|  | 16 | 765 | 753 | -1.6% |
| f15_9bits | 32 | 6039 | 6064 | +0.4% |
|  | 64 | 3303 | 3267 | -1.1% |
|  | 128 | 2051 | 1893 | -7.7% |

## 4.0 Analysis of the Simulator

In this section we use the analytical model to analyze the scalability of the simulator as well as to determine the effectiveness of other mechanisms such as the dynamic linked data structure. Figure 10 shows the effect of the dynamic data structure on the sequential simulation time for both the T3E and the SP2. To calculate the execution times in Figure 10, we assumed that the entire simulation fits in the main memory of a single processor. This allows us to isolate the effect of the dy-namic data structure. As Figure 10 shows there is a substantial gain for using a few link bits, i.e. three to seven, and the overhead of traversing the structure does not become a factor until the number of link bits approaches the total number of QU-bits. For the Mult2 and Mult3 benchmarks the use of seven link bits causes a sudden drop in the simulation time. This is because there are bits in these benchmarks which are not used until later in the calculation, and there is a substantial savings only if we can allocate these bits at the link level. A simulation of the f15_9bit benchmark shows the biggest decrease in time, where the use of five link QU-bits results in a simulation time which is one fourth that of a flat structure.

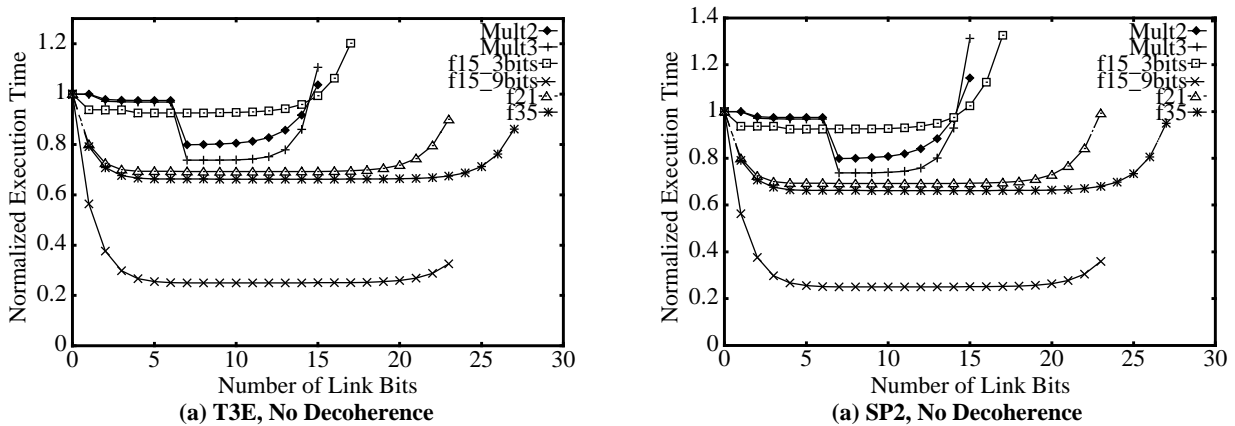

**FIGURE 10. The effect of the dynamic vector structure on the simulation time**

Figure 11 shows the speedup of the parallel simulator as a function of the number of processors for the T3E and the SP2. We compare the speedup for each benchmark to the ideal linear speedup for the specified number of processors. Each simulation point uses the number of link bits which gives the minimum simulation time. The figure shows that as the problem size becomes larger we can effectively utilize more processors. For each sized problem the speedup is close to ideal up to a certain point where it then tails off. For the smallest problem, the Mult2 benchmark, the simulation time on the T3E shown in Figure 11(a) decreases from 288 seconds, running on a single processor, to 22 seconds, running on 16 processors. The largest problem, the f35 benchmark, exhibits near linear speedup for as many as 1024 processors. This simulation on the T3E without decoherence would take close to 9 hours.

As Figure 11 shows the speedup for simulations which include decoherence is higher than simulations which do not include it. For example on the T3E, simulation of the f21 problem using 1024 processors has a speedup of 358 if we do not include decoherence, but a simulation with decoherence exhibits a speedup of 563. This is because simulating decoherence triples the total execution time without changing the reorganization time. This effectively reduces the percentage of time spent reorganizing the data. Figure 11 also shows that the speedup on the T3E and SP2 is very similar for 32 or fewer processors. For more than 32 processors, the T3E exhibits better speedup because its communication overhead is lower.

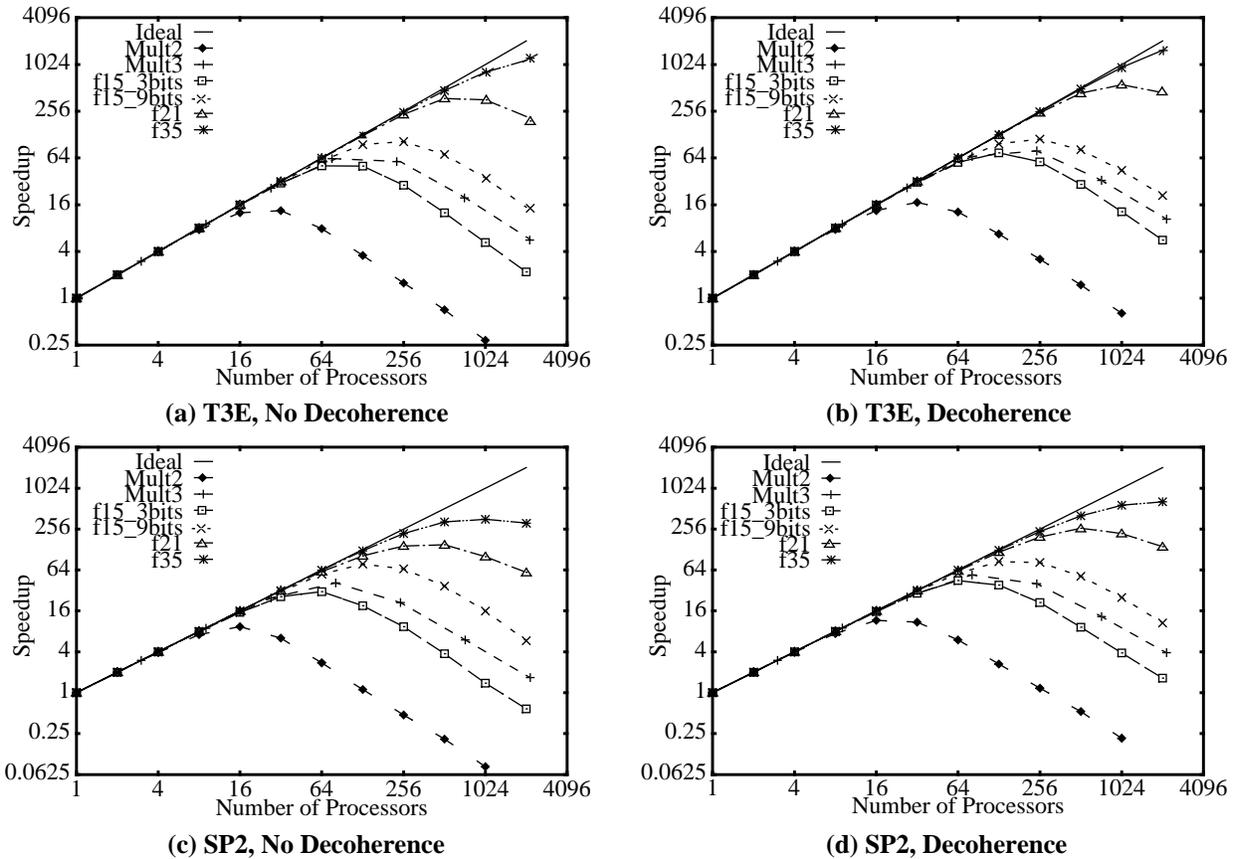

**FIGURE 11. Speedup of the Parallel Simulator**

Figure 12 shows the significance of overhead for simulations of the three largest benchmarks on the T3E. Figure 12 breaks the total simulation time into three parts, the time needed to perform the transformations on the state data, the time spent traversing the link structure and the communication time. For the f15_9bits benchmark, computation represents over 90% of the execution time for up to 64 processors. For a simulation without decoherence employing more than 64 processors, the overhead of traversing the link structure is slightly greater than the communication time.

The link overhead becomes more significant in simulations which include decoherence. The added decoherence operations add to the total link traversal time while the reorganization time remains constant. Decoherence simulations also require an accumulation operation amongst all processors after every operation, but this does not add any appreciable overhead.

For the f21 benchmark the link or communication overhead does not become significant until more than 128 processors are employed. For the f35 benchmark overhead becomes significant even later, where the simulation time for 256 processors consists mostly of calculation. In contrast to the f15_9bit benchmark, in both the f21 and the f35 benchmarks the communication time is the largest source of overhead. This is due to the fact that in the f21 and f35 benchmarks the amount of parallelism is fairly even throughout the simulation. In the f15_9bit benchmark, there is a lot of parallelism in the second half of the calculation, but less in the first half. This causes the simulator to use more link bits to expose additional parallelism, thereby increasing the link traversal time.

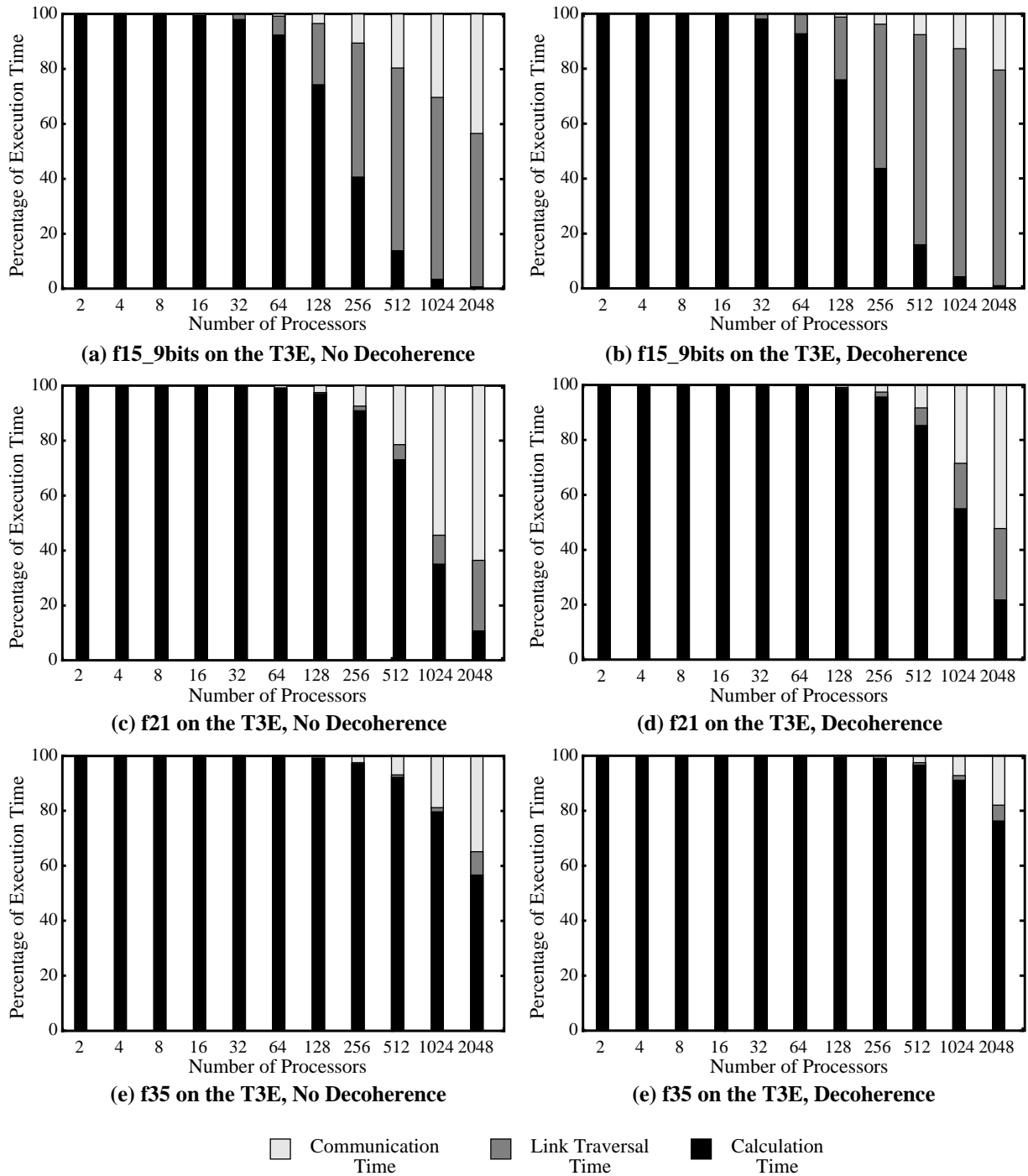

**FIGURE 12. Percentage of execution time spent doing calculation, traversing the link structure and communicating data.**

Figure 13 shows how the number of link bits affects the parallel simulation time. To exploit a certain level of parallelism the simulator requires a minimum number of link bits. Adding more link bits increases the bits which can be parallelized across, thereby decreasing the total number of re-organization steps. But as Figure 13 shows using more than the minimum number of link bits does not have a significant affect on the simulation time. Also when the number of link bits approaches the total number of bits, the execution time increases dramatically because of the added link and communication overhead. Figure 13 also shows that the link overhead becomes more significant as we employ more processors. This is due to the fact that each processor always maintains the complete link structure. Therefore the link overhead on a single processor is always the same, and it becomes a larger factor in the total simulation time as we use more processors.

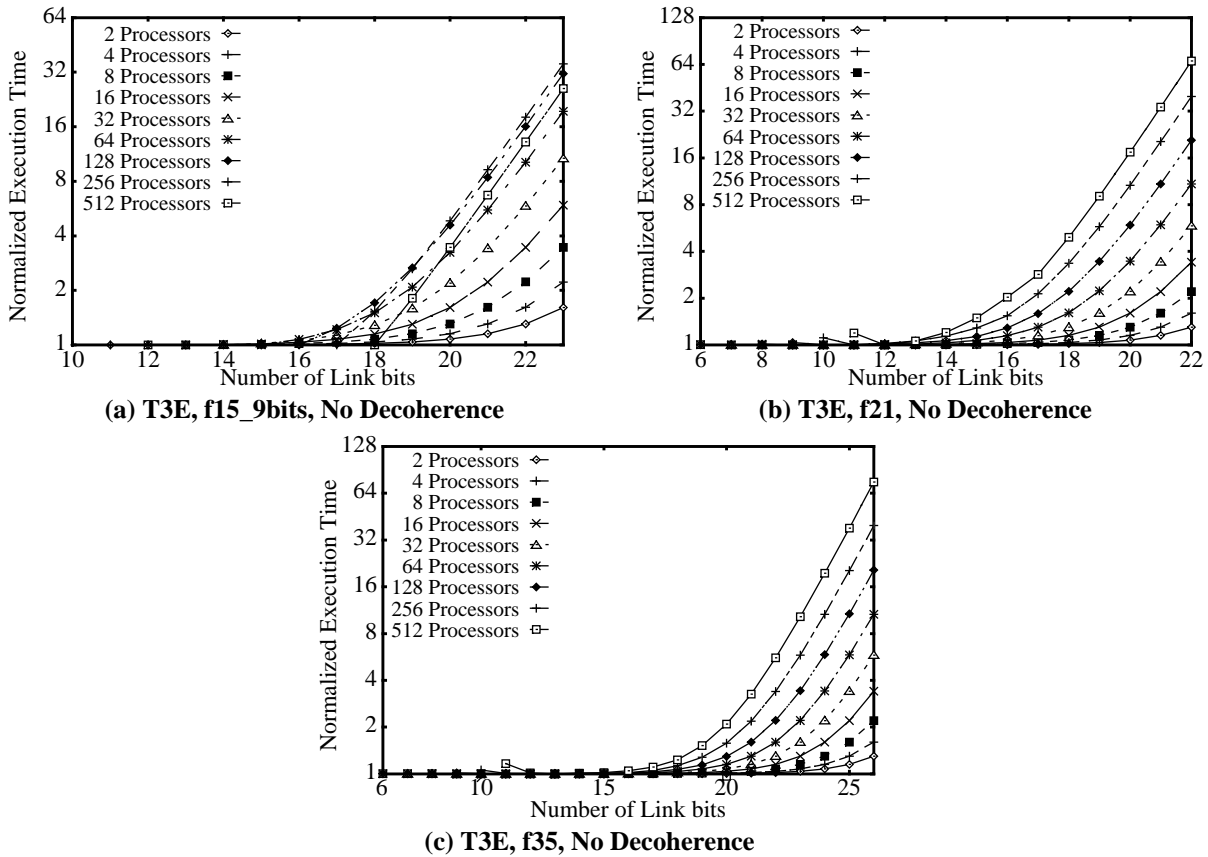

**(a) T3E, f15_9bits, No Decoherence**

**(b) T3E, f21, No Decoherence**

**(c) T3E, f35, No Decoherence**

**FIGURE 13. Effect of the number of links bits on the parallel simulation time**

# 5.0 Conclusion

In this paper we described a parallel simulator which is useful for accessing the feasibility of implementing a quantum computer. Quantum computing is a new field and therefore our simulator is one of the only tools of its kind. We find that our parallel simulator is very scalable. For many problems the simulator exhibits close to ideal speedup for as many as 256 processors. The maximum speedup scales with the size of the problem where larger problems can effectively utilize more processors than smaller problems. This allows us to simulate large problems using a supercomputer in the same amount of time it takes to simulate smaller problems on a single processor.

Simulating a quantum computer is a difficult problem because to faithfully model a quantum computer a simulator must allocate an exponential number of states and perform an exponential number of operations. Because of these requirements quantum computer simulation is a natural candidate for parallel processing. The simulator distributes the state space of the quantum computer across multiple processors and each processor performs a portion of the total calculation.

To analyze the simulator we derived an analytical model of execution time. This model allows us to determine the feasibility of simulating various sized problems. We can also use the analytical model to determine the most efficient allocation of the quantum computer's state space. Because our simulator performs quantum computer operations using matrix calculations, other types of problems which use matrix operations may also benefit from the techniques described in this paper.